\documentclass[aps,prb,twocolumn,groupedaddress, showpacs,showkeys]{revtex4}
\usepackage{graphicx,psfrag}
\usepackage{mathptmx}
\usepackage{epstopdf}
\usepackage{hyperref}
\usepackage{amsmath}
\usepackage{color}
\usepackage[latin1]{inputenc}

\begin{document}


\title{Band gap bowing of binary alloys:  Experimental results compared to theoretical tight-binding supercell calculations for Cd$_{x}$Zn$_{1-x}$Se}

\author{Daniel Mourad}
\email{dmourad@itp.uni-bremen.de}
\affiliation{Institute for Theoretical Physics, University of
Bremen, 28359 Bremen, Germany}
\author{Carsten Kruse}
\affiliation{Institute of Solid State Physics, University of
Bremen, 28359 Bremen, Germany}
\author{Sebastian Klembt}
\affiliation{Institute of Solid State Physics, University of
Bremen, 28359 Bremen, Germany}
\author{Reiner Retzlaff}
\affiliation{Institute of Solid State Physics, University of
Bremen, 28359 Bremen, Germany}
\author{Mariuca Gartner}
\affiliation{Institute of Physical Chemistry ``Ilie Murgulescu'' of the Romanian Academy, Spl.\,Independentei 202, 060021 Bucharest, Romania}
\author{Gerd Czycholl}
\affiliation{Institute for Theoretical Physics, University of
Bremen, 28359 Bremen, Germany}
\author{Detlef Hommel}
\affiliation{Institute of Solid State Physics, University of
Bremen, 28359 Bremen, Germany}
\author{Mihai Anastasescu}
\affiliation{Institute of Physical Chemistry ``Ilie Murgulescu'' of the Romanian Academy, Spl.\,Independentei 202, 060021 Bucharest, Romania}



\date{\today}
\begin{abstract}
Compound semiconductor  alloys of the type A$_x$B$_{1-x}$C  find widespread applications as their electronic bulk band gap varies continuously with $x$, and therefore a tayloring of the energy gap is possible by variation of the concentration. We model the electronic properties of such semiconductor alloys by a multiband $sp^3$ tight-binding model on a finite ensemble of supercells and determine the band gap of the alloy. This treatment allows for an intrinsic reproduction of band bowing effects as a function of the concentration $x$ and is exact in the alloy-induced disorder. In the present paper, we concentrate on bulk Cd$_x$Zn$_{1-x}$Se as a well-defined model system and give a careful analysis on the proper choice of the basis set and supercell size, as well as on the necessary number of realizations. The results are compared to experimental results obtained from ellipsometric measurements of Cd$_x$Zn$_{1-x}$Se layers prepared by molecular beam epitaxy (MBE) and photoluminescence (PL) measurements on catalytically grown Cd$_x$Zn$_{1-x}$Se nanowires reported in the literature.
\end{abstract}


\maketitle


\section{Introduction\label{sec:intro}}
For a wide class of compound semiconductor materials AC and BC, alloys of the type A$_x$B$_{1-x}$C can be realized. These substitutional alloys find a large variety of applications, in particular in nanoelectronic structures like quantum wells, quantum wires and quantum dots. Since the bulk band gap varies continuously with $x$,  a tayloring of the energy gap is possible by variation of the concentration. Most of the II-VI and III-V compound semiconductors, e.\,g.\,CdSe, ZnSe, CdS, CdTe, GaAs or InAs possess a direct band gap, covering the entire visible spectrum, which makes them particularly suitable in optoelectronic and especially photonic devices.

Many of these materials show a pronounced bowing behaviour of the bulk band gap  as a function of the concentration $x$. The deviation from linear behaviour can often be described using a single bowing parameter $b$. The concentration dependent energy gap $E_{\text{g}}(x)$ of the mixed crystal is then given by
\begin{equation} \label{eq:bandgapbowing}
 E_{\text{g}}(x) = x \, E_{\text{AC}} + (1-x) \, E_{\text{BC}} - x \, (1-x) \, b,
\end{equation}
where $E_{\text{AC}}$ and $E_{\text{BC}}$ are the corresponding gaps of the pure AC and  BC material. When the pure materials AC and BC have considerably different lattice constants, also a non-parabolic dependence of the band gap on $x$ is observed.~\cite{richardson_dielectric_1973} Nevertheless, a fit of  experimental band gap data  by the ansatz (\ref{eq:bandgapbowing}) is still common and adequately possible  for materials like Cd$_x$Zn$_{1-x}$Se, Al$_x$Ga$_{1-x}$As, In$_x$Ga$_{1-x}$As and many more.

The corresponding literature values for the parameter $b$ usually show a large variety, depending on the theoretical or experimental approach used. In the present paper we will concentrate on bulk Cd$_x$Zn$_{1-x}$Se, where a broad range of values between $b=0$  and $b=1.26$  eV is reported.~\cite{ammar_structural_2001, venugopal_photoluminescence_2006, tit_absence_2009, gupta_optical_1995} From the experimental point of view, this discrepancy  considerably originates from the fact that the stable phase of pure CdSe is the wurtzite phase, while pure ZnSe will crystallize in the cubic zincblende modification. Also, the experimental realization has to deal with phase separation effects and the miscibility gap of the alloys. This makes it difficult to grow homogeneously alloyed Cd$_x$Zn$_{1-x}$Se bulk samples over the whole concentration range. These problems carry over directly to the theoretical side, as the experimental determination of material properties and hence input parameters for calculations involving zincblende CdSe is very difficult and will often depend on the substrate on which the sample was grown. 

In the past, a broad spectrum of different computational methods has  been applied to calculate the band gaps of alloyed II-VI and III-V bulk systems, from highly-sophisticated methods like empirical pseudopotential \cite{bernard_optical_1986} or DFT + LDA \cite{bernard_electronic_1987} models, over empirical tight-binding models (ETBM) \cite{tit_origins_2009, tit_absence_2009} to simple two band models\cite{van_vechten_electronic_1970}. Within the ETBM, one common approach is to use a plane-wave basis set and implicitly map the alloyed  and therefore disordered systems onto an effective translationally invariant system by either using a simple linear interpolation of relevant parameters (virtual crystal approximation, VCA) or more sophisticated Green function methods (coherent potential approximation, CPA). Because the VCA is known to vastly underestimate the band gap bowing in most cases,~\cite{boykin_approximate_2007} and the standard CPA approaches also fail to give quantitatively accurate results,~\cite{hass_electronic_1983, laufer_tight-binding_1987} tight-binding (TB) methods that use a localized basis set on a finite supercell\cite{boykin_practical_2005} are more promising nowadays. These models, albeit computationally costful, are able to reproduce bowing effects and are exact in the disorder.

To calculate the electronic properties of such semiconductor alloys, we start from multiband $sp^3$ tight-binding models for the pure bulk semiconductor materials AC and BC and then perform calculations on a finite ensemble of supercells.  We will make  recommendations for the cell size as well as for the necessary number of realizations, based on the convergence  of the bowing parameter. These recommendations will then be verified by comparison to band gaps as determined from photoluminescence data for catalytically grown Cd$_{x}$Zn$_{1-x}$Se nanowires by Venugopal \textit{et al}.~\cite{venugopal_photoluminescence_2006} 
We also present new experimental results obtained for Cd$_{x}$Zn$_{1-x}$Se films grown by MBE, for which the band gap is determined by an analysis of absorption data. A good agreement between these experimental results and the theoretical results of the TB supercell approach is obtained.
\section{Theory \label{sec:theory}}

\subsection{Empirical tight-binding model for pure bulk semiconductors: The effective bond-orbital model \label{subsec:ETBMbulk}}


Linear combinations of atomic orbitals can be used as ansatz for the Wannier functions that are localized within the unit cell of the Bravais lattice. From these Wannier functions, the itinerant Bloch functions can be determined by a unitary transformation. Because neither the atomic states nor the Wannier functions are explicitly used in an ETBM, one can directly assume a basis of Wannier-like functions, which is the basic idea of the so-called effective bond-orbital model (EBOM). This resolution on the scale of unit cells is especially convenient for the case of binary semiconductor compounds of the type A$_x$B$_{1-x}$C, as each unit cell can either be occupied by the AC- or BC-material, respectively.

In order to adequately reproduce one $s$-like conduction band (CB) and three $p$-like valence bands, the heavy hole (HH), light hole (LH) and split-off (SO) band, we use a localized $sp^3$ basis per spin direction:
\begin{equation}\label{eq:sp3basis}
\left| \mathbf{R}, \alpha \right\rangle, \quad \alpha \in \left\lbrace s\uparrow,p_x\uparrow,p_y\uparrow,p_z\uparrow,s\downarrow,p_x\downarrow,p_y\downarrow,p_z\downarrow \right\rbrace.
\end{equation}
Here $\mathbf{R}$ labels the $N$ sites of the fcc lattice, which is the underlying Bravais lattice of the zincblende crystal structure.

The TB matrix elements of the bulk Hamiltonian $H^{\text{bulk}}$  are then given by
\begin{equation}\label{eq:EBOMme}
E_{\alpha \alpha'}^{\mathbf{R} \mathbf{R'}} = \left\langle \mathbf{R}, \alpha \right| H^{\text{bulk}} \left| \mathbf{R'}, \alpha' \right\rangle.
\end{equation}
When restricting the non-vanishing matrix elements, Eq. (\ref{eq:EBOMme}), up to second nearest neighbors, a one-to-one correspondence between the TB matrix elements and a set of material parameters can be determined. In the present work, we use the parametrization scheme of Loehr,~\cite{loehr_improved_1994} which fits the bulk band structure to a set of material parameters, namely the bulk band gap $E_{\text{g}}$, the conduction band effective mass $m_e$, the spin-orbit splitting $\Delta_{\text{so}}$, the conventional lattice constant $a$ and additionally to the X-point energies of the conduction
band $X_{\text{1c}}$, the  HH and LH bands $X_{\text{5v}}$ (which are degenerate at the Brillouin zone boundary), and the SO band $X_{\text{3v}}$. We use a set of material parameters from Refs.\,\onlinecite{kim_optical_1994, hlscher_investigation_1985}, augmented by $X$-point energies from Ref.\,\onlinecite{blachnik_numerical_1999}. This Hamiltonian has already been used succesfully for bulk as well as nanostructure calculations for several material systems.~ For detailed comparisons with other well-established models, we refer to Refs. \onlinecite{marquardt_comparison_2008,  schulz_multiband_2009}. Our specific choice of values for the material parameters for CdSe and ZnSe has been proven to be reliable in calculations for pure quantum dots \cite{schulz_tight-binding_2005} as well as for mixed Cd$_{x}$Zn$_{1-x}$Se colloidal nanocrystals.~\cite{mourad__2010}

\subsection{Treatment of alloying in the effective bond-orbital model\label{subsec:alloying}}

The usual approach for translationally invariant systems like pure bulk crystals is the construction of a plane-wave basis set from Eq. (\ref{eq:sp3basis}), which then fulfills the Bloch condition. For the Cd$_{x}$Zn$_{1-x}$Se alloy system, we will now assume uncorrelated substitutional disorder and hence lose the translational symmetry. In analogy to calculations for zero-dimensional nanostructures, ~\cite{marquardt_comparison_2008, schulz_multiband_2009, mourad_multiband_2010} we will use a finite supercell with periodic boundary conditions, with the localized basis, Eq.\,(\ref{eq:sp3basis}), on each lattice site. Each primitive cell will either be occupied by the diatomic CdSe or ZnSe basis, where the probability for the respective species is directly given by the concentrations $x$ and $1-x$. Accordingly, we will use the matrix elements, Eq. (\ref{eq:EBOMme}) of the pure CdSe or ZnSe material for the corresponding lattice sites and use a linear interpolation for hopping matrix elements between unit cells of different material. Due to the resolution on the Bravais lattice, all influences which stem from effects on a smaller length scale, like the difference in the Cd--Se and Zn--Se bond lengths, are  absorbed into the values of the corresponding  TB matrix elements between the ``effective`` orbitals.
Furthermore, we incorporate an appropriate valence band offset (VBO) between the two materials by shifting the respective site-diagonal matrix elements by the value $\Delta E_{\text{vb}} = 0.22$ eV, as established in Ref.\,\onlinecite{schulz_tight-binding_2005}. Although this treatment of the intersite hopping and the VBO has merely been well-proven for interfaces in quantum dot systems,~\cite{mourad_multiband_2010, marquardt_comparison_2008, schulz_multiband_2009} our results for the bowing have turned out to be insensitive not only to the choice of the interface treatment but also to the specific value of $\Delta E_\mathrm{vb}$ (not shown).

\begin{figure}
	\includegraphics[width = \linewidth]{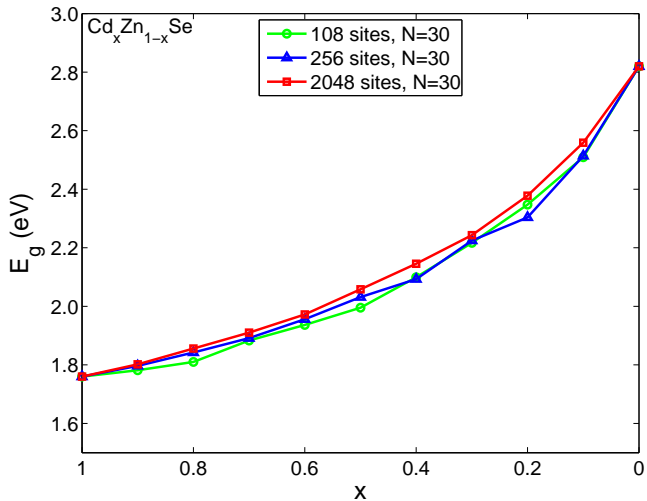}
\caption{(Color online) Dependance of the bulk band gap $E_{\text{g}}$ on the concentration $x$ of Cd$_{x}$Zn$_{1-x}$Se for different supercell sizes. Each lattice site corresponds to one unit cell and hence two atoms, while $N=30$ gives the number of distinct microscopic configurations.}
\label{fig:sizedep}
\end{figure}

The numerical diagonalization of the corresponding Hamiltonian for each concentration and microscopic configuration gives the  density of states (DOS) of the finite ensemble, from which the band gap for each concentration can be determined as the difference between the overall lowest conduction band and highest valence band energy.  Fig.\,\ref{fig:sizedep} shows the corresponding band gap over the concentration in dependence on the number of lattice sites (for fixed number of realizations), where each lattice site corresponds to one unit cell and thus two atoms. Obviously, too small supercells produce an artificially high band gap bowing even with periodic boundary conditions, and this finite size effect cannot be compensated for by a higher number of realizations (not shown here). This could be attributed to clustering effects, as a local accumulation of the small band gap material will lead to an overall smaller band gap from the DOS. Our studies reveal that statistically relevant convergence for the bowing $b$ is safely reached with 4000 lattice sites (8000 atoms) and $N=50$ realizations when seeking for a resolution of 10 meV, which corresponds to the last digit of the input energies. With a cubiform supercell, which does not artificially confine the degrees of freedom of the microscopic disorder, this corresponds to an edge length of $10 \, a$. Of course, these large supercells and high number of configurations require large computational costs.

From our results, we cautiously conclude that ETBM calculations which use a lesser number of atoms on rectangular supercells (like Ref.~\onlinecite{boykin_approximate_2007}, where an approximate band structure is reconstructed with a $40 \, a \times 2 \, a \times 2 \,a$ supercell, i.\,e.\,1280 atoms) could therefore  overestimate the bowing of the alloy band gap. We performed first test calculations for other material systems (e.\,g.\, Ga$_{x}$Al$_{1-x}$N) as well and the convergence criterion of 4000 sites and 50 realizations remained still valid.

Finally, we want to dicuss the distributions of the band gap values from the different configurations. In theory, the difference between the lowest overall conduction band energy and the highest overall valence band energy corresponds to the band gap $E_{\text{g}}(x)$ of the mixed A$_{x}$B$_{1-x}$C crystal, as approximated by the finite ensemble. This is also the usual way to determine band gaps in the CPA,~\cite{hass_electronic_1983, laufer_tight-binding_1987} which can reproduce broadening effects due to disorder.  When referring to the energetic positions obtained from  PL peak positions (other effects like the Stokes shift left aside for the moment), one should in principle compare to the ensemble-average $E_{\text{g}}^{\text{av}}(x) = 1/N \sum_{i}E_{\text{g}}^i(x)$ of the band gap, because the effects of configurational and concentrational disorder, amongst others, are already absorbed into the finite emission linewidth. $E_{\text{g}}^i$ is here the difference between the lowest conduction band state and the highest valence band state for the $i^{\text{th}}$ microscopic configuration.

\begin{figure}
	\includegraphics[width = \linewidth]{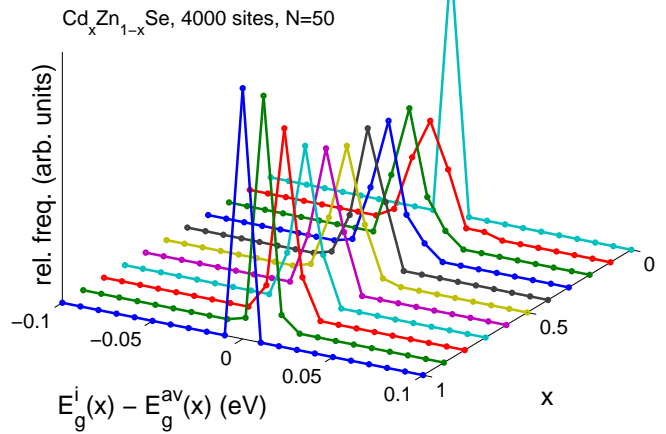}
\caption{(Color online) Distribution of the deviation of the band gap $E_{\text{g}}^i(x)$, as calculated from the $i^{\text{th}}$ microstate, from the average value $E_{\text{g}}^{\text{av}}(x)$ for each concentration $x$. The supercell contains 4000 lattice sites and $N=50$ realizations per concentration were chosen.}
\label{fig:distr}
\end{figure}

In Fig.\,\ref{fig:distr}, we have depicted the relative frequency of the deviation of  $E_{\text{g}}^i(x)$ from the average value $E_{\text{g}}^{\text{av}}(x)$ for Cd$_{x}$Zn$_{1-x}$Se. As discussed above, we used 50 distinct microscopic configurations and the supercell was chosen to contain 4000 sites in order to reach convergence. The distribution of the $E_{\text{g}}^i(x)$ is quite narrow for most concentrations $0 < x < 1$ and, of course, $\delta$-like for the pure system ($x=0$ and $x=1$), where all microstates are identical. The biggest deviations, approximately 20 meV, between  $E_{\text{g}}^i(x)$ and $E_{\text{g}}^{\text{av}}(x)$ occur for $x=0.1$. Further studies (not shown) reveal that only the conduction band edge is broadened due to the disorder, while the valence band edge is $\delta$-like over the whole concentration range within our finite resolution, which is in agreement with CPA calculations for other II-VI alloys.~\cite{hass_electronic_1983} The overall band gap of the finite ensemble is hence given by
\begin{equation}
E_{\text{g}}(x) = \operatorname{min}\left\{ E_{\text{g}}^i(x) \right\}.
\end{equation} 
In the present work, we used  $E_{\text{g}}^{\text{av}}$ for the comparison to PL data and the overall band gap of the ensemble DOS $E_{\text{g}}$ for the absorption experiments.  Nevertheless, the distributions are so narrow that the bowing parameters calculated in this work only change within the error boundaries when switching between both approaches, provided that the above mentioned convergence criteria are fulfilled.

\section{Comparison to experimental results \label{sec:experimental}}

\subsection{Comparison to PL on bulk-like nanowires\label{subsec:nanowires}}

Venugopal \textit{et al.} performed photoluminescence (PL) and Raman-scattering measurements  on catalytically grown Cd$_{x}$Zn$_{1-x}$Se nanowires (60 - 150 nm in diameter and several tens of micrometers in length) in Ref.\,\onlinecite{venugopal_photoluminescence_2006}. From these data, they determined the band gap as a function of the concentration $x$ and the corresponding bowing parameter. Our TB parameter set was chosen so that their respective experimental band gaps at room temperature for the pure systems, $x=0$ and $x=1$, are reproduced in our calculations. As opposed to the films discussed in Sec. \ref{subsec:thinfilms}, the Cd-richer nanowires  still tend to crystallize in the wurtzite phase. Regardless, the EBOM results  from the localized basis set, Eq. (\ref{eq:sp3basis}), for the band gap will remain valid~\cite{von_grnberg_energy_1997} when the only slightly larger energy gap of hexagonal CdSe is used in the calculations, especially as the experimental data does not show any discontinuity.

\begin{figure}
	\includegraphics[width = \linewidth]{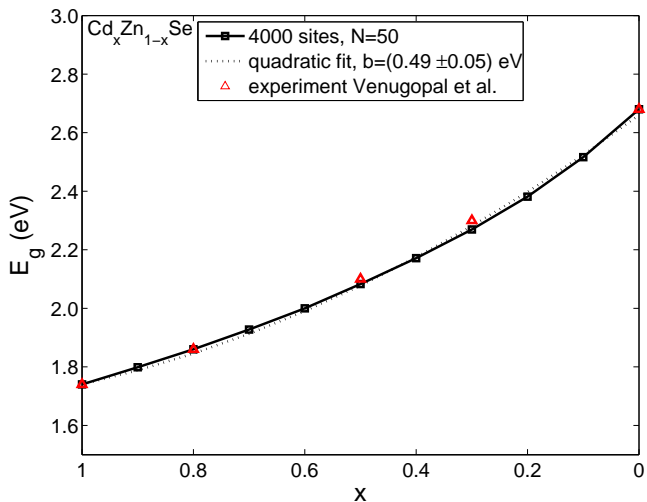}
\caption{(Color online) Dependance of the  band gap $E_{\text{g}}$ on the concentration $x$ of Cd$_{x}$Zn$_{1-x}$Se bulk-like nanowires. The black squares give the ETBM ensemble results for 4000 lattice sites and 50 realizations, while the red triangles give the results of PL measurements on bulk-like Cd$_{x}$Zn$_{1-x}$Se nanowires by Venugopal \textit{et al}.~\cite{venugopal_photoluminescence_2006} Although only fitted to $x=1$ and $x=0$, the ETBM results show a very good agreement over the whole concentration range.}
\label{fig:nanowires}
\end{figure}

The resulting curve and the PL results are given in Fig.~\ref{fig:nanowires} and show a very good agreement over the whole concentration range.  A quadratic fit yields a bowing parameter of $b=(0.49 \pm 0.05)$ eV, which is in excellent agreement with the value of $b=(0.45\pm 0.02)$ eV from Ref.\,~\onlinecite{venugopal_photoluminescence_2006}. 

\subsection{Comparison to epitaxial films\label{subsec:thinfilms}}

The samples have been prepared by molecular beam epitaxy (MBE) on a GaAs (100) substrate at a temperature of 280$^{\circ}$C. The binary ZnSe and CdSe layers have been grown directly on the substrate, while for the ternary alloys first a ZnSe buffer layer of 20 nm thickness has been deposited, followed by the Cd$_x$Zn$_{1-x}$Se layer. In order to realize different Cd concentrations of the ternary compound, the operating temperature of the Cd Knudsen cell was varied from 226~-~240$^{\circ}$C during deposition. Special care was taken to ensure stoichiometric or slightly Se-rich growth conditions for each case by adjusting the respective flux using both a Se Knudsen cell and a valved cracker source at the same time. A valved cracker source offers the advantage of rapid variations in elemental flux due to the availability of a motorized valve. The stoichiometry has been controlled in-situ employing a reflection high-energy electron-diffraction (RHEED) system. A 2$\times$1 reconstruction indicates group VI-rich growth, while a c(2$\times$2) reconstruction can be assigned to metal-rich (i.\,e.\,Zn and Cd) conditions. Stoichiometry is achieved when a superposition of both reconstructions can be identified within the RHEED pattern. The growth rate of the Cd$_x$Zn$_{1-x}$Se alloy depends sensitively on the II/VI flux ratio. Since the deposition time for Cd$_x$Zn$_{1-x}$Se was kept constant (2 hours) for the relevant samples, the thickness of the different Cd$_x$Zn$_{1-x}$Se layers show a significant variation. However, this does not influence the accurate determination of the band gap as discussed below. 
\begin{figure}
	\includegraphics[width = \linewidth]{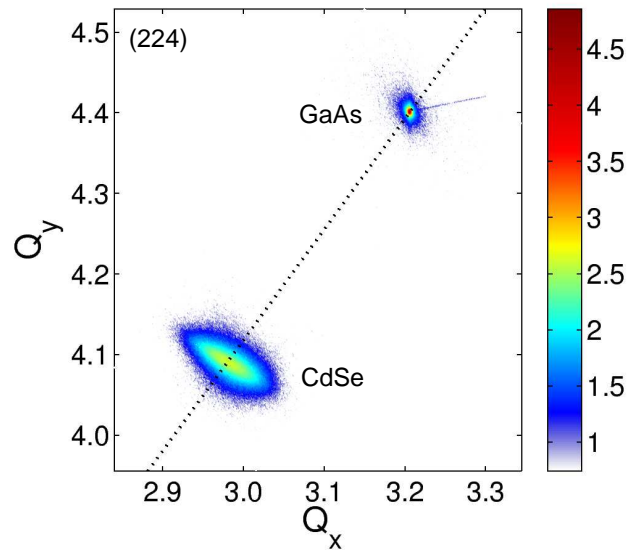}
\caption{(Color online) Reciprocal space map of the (224) reflex for the binary CdSe sample, measured by high-resolution X-ray diffraction. The CdSe layer is fully relaxed, since it lies on a straight line that passes through the origin of the graph.}
\label{fig:recspacemap}
\end{figure}
The Cd content and the respective strain state of the Cd$_x$Zn$_{1-x}$Se layers has been determined by the evaluation of reciprocal space maps (RSM) in the vicinity of the (224) reflection recorded using high-resolution X-ray diffraction (HRXRD). The RSM measured for the binary CdSe sample ($x=1$) is examplarily shown in Fig.\,\ref{fig:recspacemap}. The peak of the CdSe layer is located on a straight line that passes through the origin of the graph, indicating that the layer is fully relaxed and therefore unstrained. The same result is found for the other Cd$_x$Zn$_{1-x}$Se films. Merely the layer of the binary ZnSe sample ($x=0$) is not fully relaxed on the GaAs substrate and therefore remains compressively strained. The absence of strain is especially important for the  TB modelling for the system under consideration, as it does not have to include the effect of strain and piezoelectricity on the electronic states, as it is often the case in calculations for quantum dot systems.~\cite{schulz_electronic_2006, schulz_spin-orbit_2008} 

In order to evaluate the optical constants, the band gap and the thickness of the Cd$_x$Zn$_{1-x}$Se films, optical measurements were carried out using a J.\,A.\,Woollam variable-angle spectroscopic ellipsometry (VASE) setup.~\cite{aschenbrenner__2010} 
The spectra cover the wavelength range from 193 nm to 1700 nm (6.42 eV - 0.7 eV) with wavelength steps of 2 nm. To get more accurate data, spectra for three different angles of incidence (65$^{\circ}$, 70$^{\circ}$  and 75$^{\circ}$) were registered and analyzed for each sample. All measurements were perfomed at room temperature. Beside the GaAs (100) substrate, three layers were taken into account, namely the ZnSe buffer, the Cd$_x$Zn$_{1-x}$Se-layer and a thin additional layer to include the surface roughness into the model. The latter one was simulated by a Bruggeman effective medium approximation (EMA) using 50\% voids in a matrix of the II-VI semiconductor material. The optical constants were derived from the curve fits using a model which combines a Herzinger-Johs ``P-semi oscillator'' (four connected functions to model each peak)\cite{herzinger_parametric_2010}
with four Gauss oscillators. 

\begin{figure}
	\includegraphics[width = \linewidth]{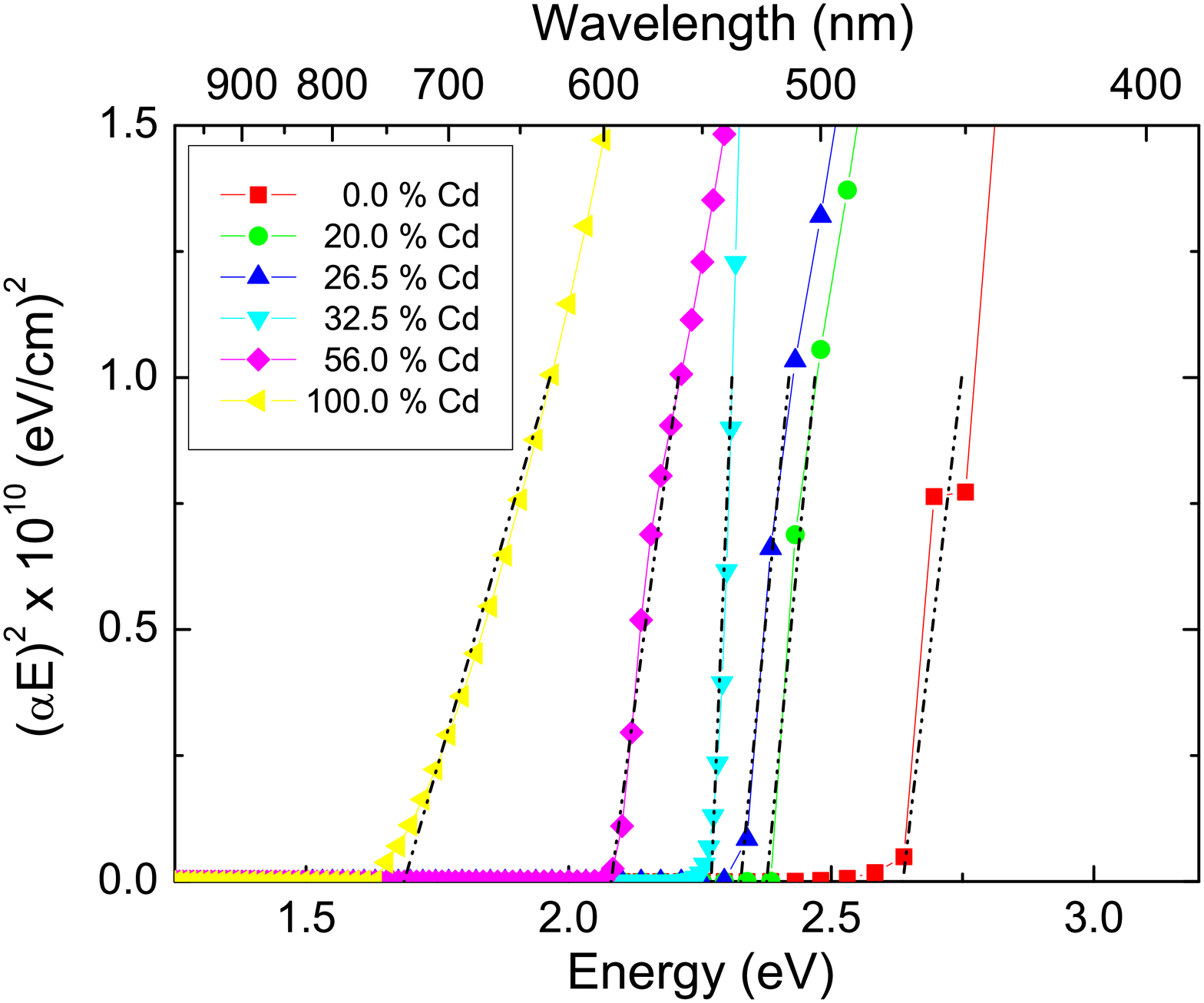}
\caption{(Color online) Absorption curves of Cd$_x$Zn$_{1-x}$Se layers in dependance of the Cd content. The black dashed lines indicate the tangent used for determination of the bandgap energy $E_{\text{g}}$.}
\label{fig:abs_edge}
\end{figure}

The band gap  $E_{\text{g}}$ of the Cd$_x$Zn$_{1-x}$Se layers was obtained using the formula\cite{pankove_optical_1975}
\begin{equation}
(\alpha E)^2 \propto  (E-E_{\text{g}}),
\end{equation}
and the equation $\alpha(E)=4\pi\kappa(E)/\lambda$ for the attenuation coefficient $\alpha$, where $\kappa(E)$ denotes the extinction coefficent, $E$ the energy  and $\lambda$ the wavelength of the incident light. The spectral extinction coefficent $\kappa(E)$ is determined by the spectroscopic ellipsometry measurement of the samples and $E_{\text{g}}$ is obtained graphically by evaluating the intersection of the tangent of $(\alpha E)^2$ with the energy axis (see Fig.\,\ref{fig:abs_edge}). A comprehensive overview of the relevant experimental parameters for the samples is given in Table~\ref{tab:table_samples}.

\begin{table}
\caption{Selected parameters of the Cd$_x$Zn$_{1-x}$Se films determined by spectroscopic ellipsometry and high-resolution X-ray diffraction.}
\label{tab:table_samples}
\begin{ruledtabular}
\begin{tabular}{llcc}
Cd content (\%) &thickness~(nm) &roughness~(nm) &$E_{\text{g}}$~(eV)\\
\hline
0.0    & 281.2  & 7.1 & 2.64~$\pm~0.02$\\
20.0   & 1396.7 & 7.8 & 2.37~$\pm~0.03$\\
26.5 & 1490.3 & 6.9 & 2.30~$\pm~0.02$\\
32.5 & 1821.0 & 7.3 & 2.25~$\pm~0.03$\\
56.0   & 1170.4 & 4.4 & 2.07~$\pm~0.02$\\
100.0  & 304.0  & 2.6 & 1.69~$\pm~0.02$\\
\end{tabular}
\end{ruledtabular}
\end{table}

For the theoretical calculations, we determined the TB parameters of the pure systems (i.\,e.\, for $x=1$ and $x=0$) from standard band structure data  in the literature,~\cite{_cubic_2004, blachnik_numerical_1999} valid at room temperature, as the input parameters for the pure cubic CdSe were determined under the same experimental conditions and on the same substrate as in the present measurements.
\begin{figure}
	\includegraphics[width = \linewidth]{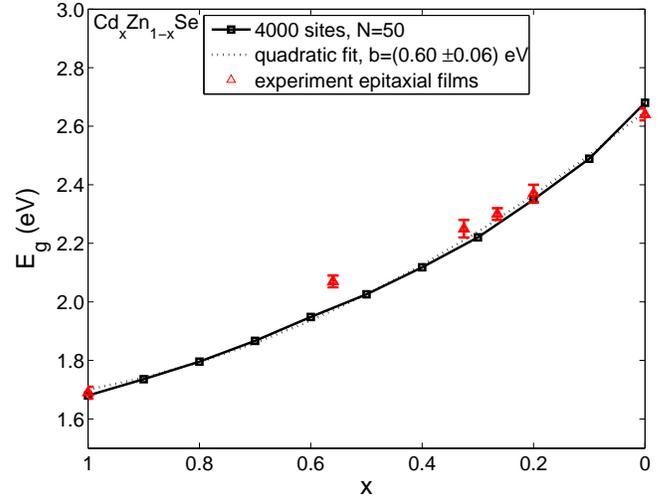}
\caption{(Color online) Dependence of the bulk band gap $E_{\text{g}}$ on the concentration $x$ of Cd$_{x}$Zn$_{1-x}$Se epitaxial films. The black squares again give the ETBM ensemble results, while the red triangles give the results as obtained from Fig.\,\ref{fig:abs_edge}. Although only material parameters for $x=1$ and $x=0$ from the literature were taken as input, we still have a good agreement between this specific experimental and the theoretical data, with exception of the value for $x=0.56$.}
\label{fig:thinfilms}
\end{figure}

The values for the band gap in dependence on the Cd fraction $x$ are depicted in Fig.\,\ref{fig:thinfilms} in comparison with the theoretical results. We get a good agreement within the error boundaries between the  experiment and the theoretical data, except for one data point with the concentration $x=0.56$, which clearly does not fit the general trend in the data. When this point is omitted, the experimental bowing coefficient is $b=(0.41\pm 0.09)$ eV, which lies below the theoretical result of $b=(0.60\pm0.06)$ eV from the TB calculations for the system under consideration. This discrepancy could for example be explained by the different experimental conditions, as especially the production of high-quality Cd$_{x}$Zn$_{1-x}$Se  films, but also the determination of the band gap from the absorption edges may be afflicted with a slightly higher degree of uncertainty than the shift-adjusted PL measurement of the nanowires of Ref.\,\onlinecite{venugopal_photoluminescence_2006}. For example, a look at Tab.\,\ref{tab:table_samples} reveals that the literature value  of $E_{\text{g}}=2.68$ eV for pure ZnSe at room temperature~\cite{blachnik_numerical_1999} cannot be exactly reproduced. For the sake of completeness, we would like to note that one obtains an experimental bowing value of $b=(0.6\pm 0.2)$ eV, which  fits the theoretical result within the error range, if the literature value is used for the binary ZnSe sample instead.

\subsection{Comparison of the bowing results\label{subsec:comparison}}

For the sake of comparison, the results for the bowing of Cd$_{x}$Zn$_{1-x}$Se for both experiments are summarized in Table\,\ref{tab:table_summary}. In spite of the overall good agreement, the results from the TB supercell calculations for the disordered systems ($1>x>0$) are all  slightly shifted to  lower energies in comparison to the experimental values, which result in the higher bowing parameter in the theory. 

\begin{table}
\caption{Summary of the results for the bowing parameter $b$ for Cd$_{x}$Zn$_{1-x}$Se at room temperature (all numbers given in  eV). Note that the crystal structure of the nanowires changes from zincblende to wurtzite for larger $x$.}
\label{tab:table_summary}
\begin{ruledtabular}
\begin{tabular}{lll}
 system &  experiment &  supercell ETBM  \\
\hline
nanowires  & $0.45 \pm 0.02$~\cite{venugopal_photoluminescence_2006} & $0.49 \pm 0.05$\\
epitaxial films  & $0.41 \pm 0.09$ & $0.60 \pm 0.06$\\
\end{tabular}
\end{ruledtabular}
\end{table}

When we have a closer look at the values for the bowing coefficient $b$, we can conclude that, in spite of the different experimental  conditions and therefore boundary values at $x=1$ and $x=0$, all bowing values lie within the comparatively narrow range between $b\approx0.4$ eV and $b\approx0.6$ eV. This stands in contrast to several previous studies, who reported a broad variety of values between $b=0$ and $b = 1.26$ eV,~\cite{ammar_structural_2001, venugopal_photoluminescence_2006, tit_absence_2009, gupta_optical_1995} as already mentioned in the introduction. Therefore, we recommend a critical review of literature values outside the range given in Tab.\,\ref{tab:table_summary} before using them for further purposes.

\section{Conclusion and outlook \label{sec:concl}}

In this paper, we presented a combined theoretical and experimental approach for the determination of the band gap bowing of binary compound semiconductor alloys of the type A$_x$B$_{1-x}$C. On the theoretical side, we used a multiband empirical tight-binding approach with a localized basis set on the lattice sites of a finite supercell with periodic boundary conditions. By calculating the band gap for each concentration $x$ from the density of states of a finite ensemble with different microscopic configurations, our approach is exact in the alloy-induced disorder. The input parameters are a set of material properties of the pure AC or BC system, i.\,e.\, for the cases $x=1$ and $x=0$.  Using the example of Cd$_x$Zn$_{1-x}$Se, we gave recommandations on the minimal size of the supercell and the minimal necessary number of microscopic configurations in order to determine the band gap bowing coefficient $b$ with the same degree of accuracy as the input parameters, as Fig.\,\ref{fig:sizedep} showed that the results depend strongly on these model parameters.

Furthermore, we compared our results for the band gap bowing of Cd$_x$Zn$_{1-x}$Se with two slightly distinct experimental realizations. At first, we compared our results to photoluminescence data from catalytically grown Cd$_{x}$Zn$_{1-x}$Se nanowires by Venugopal \textit{et al}. Then we used our TB model for an analysis of  absorption data from unstrained MBE-grown Cd$_{x}$Zn$_{1-x}$Se films, and also gave a detailed description of the experimental realization of this data set. In both cases, our model fits the experimental data well. For the nanowires PL data, 
even the experimental band gap bowing parameter can be reproduced within the error margin, while the measurements on the epitaxially grown  films show a slightly higher deviation from the theoretical predictions. Nevertheless, within the experimental accuracy the bulk band gaps from all but one MBE samples fit the theoretical curve well. Dependent on the specific experimental conditions, we carefully approve to narrow down the range of the bowing parameter for bulk Cd$_{x}$Zn$_{1-x}$Se at room temperature to between $b\approx0.4$ eV and $b\approx0.6$ eV. 

 In contrast to common tight-binding models, which treat the disorder within a mean-field framework, like the virtual crystal approximation, our simple TB supercell approach is able to satisfactorily reproduce the experimental findings and can thus be used to coherently calculate bowing coefficients for other material systems and growth conditions, which in addition may not be experimentally accessible over the whole concentration range.
For this purpose, our model can easily be transferred to other material systems like  Cd$_x$Zn$_{1-x}$S, Al$_x$Ga$_{1-x}$As, In$_x$Ga$_{1-x}$As, Ga$_x$Al$_{1-x}$N, and many others in order to calculate the bowing of the band gap. Also, materials with a wurtzite structure can be simulated, as only the corresponding TB Hamiltonian in the effective bond-orbital model has to be replaced by the one given in Ref.\,\onlinecite{mourad_multiband_2010}. First results indicate that our minimal criteria for the numerical convergence of the band gap bowing will still remain valid, although this will have to be tested carefully for each system under consideration, in combination with the respective experimental data obtained by means of epitaxy.

\begin{acknowledgments}

The authors would like to thank Jan-Peter Richters and Tobias Vo{\ss} for fruitful discussions. The financial support of the Deutsche Forschungsgemeinschaft (Project No. 436 RUM 113/27/0-1) is gratefully acknowledged.

\end{acknowledgments}


\appendix



\end{document}